# The known unknowns of the Hsp90 chaperone

**Laura-Marie Silbermann[1]‡, Benjamin Vermeer[2]‡, Sonja Schmid[2]\*, Katarzyna (Kasia) Tych[1]\***

[1]Groningen Biomolecular Sciences and Biotechnology Institute, University of Groningen, Nijenborgh 7, 9747 AG Groningen, the Netherlands

[2]Laboratory of Biophysics, Wageningen University & Research, Stippeneng 4, 6708 WE Wageningen, The Netherlands

‡Equal contributions.

\*Correspondence to: k.m.tych@rug.nl, schmid@nanodynlab.org.

**Abstract**

Molecular chaperones are vital proteins that maintain protein homeostasis by assisting in protein folding, activation, degradation, and stress protection. Among them, heat-shock protein 90 (Hsp90) stands out as an essential proteostasis hub in eukaryotes, chaperoning hundreds of "clients" (substrates). After decades of research, several "known unknowns" about the molecular function of Hsp90 remain unanswered, hampering rational drug design for the treatment of cancers, neurodegenerative and other diseases. We highlight three fundamental open questions, reviewing the current state of the field for each, and discuss new opportunities, including single-molecule technologies, to answer the known unknowns of the Hsp90 chaperone.

Proteins are the active workforce in the cell. Among them, chaperone proteins are responsible for maintaining *proteostasis*, i.e., sustaining a functional proteome adapted to ever-changing cellular and environmental conditions. In eukaryotic cells, the Heat shock protein 90 (Hsp90) plays a central role in proteostasis, as it provides a scaffold that binds substrate proteins (termed clients) as well as a variety of helper chaperones (known as cochaperones)[1–8]. Hsp90 can therefore be regarded as a versatile workbench[7] where cochaperones act as tools to customise the chaperone's function as needed for client recruitment[9–11], (re)folding[12,13], (de)activation[14,15], degradation, or protection from degradation[7,16]. Due to the many clients relying on its function, Hsp90 constitutes a proteostatic hub in eukaryotes, while only a handful of prokaryotic Hsp90 clients are known[10,11]. Among the hundreds of eukaryotic clients are transcription factors, signalling kinases, DNA replication proteins, cell division regulators, synapse proteins, and many more[1]. Hsp90 is highly conserved throughout prokaryotes to higher eukaryotes[17], and appears to have back-transferred from eukaryotes to their archaeal precursors through horizontal gene transfer[18]. In the absence of stress, Hsp90 constitutes 1 – 2% of the protein mass in eukaryotic cells[19] and exists in super-stoichiometric ratios (~2:1) to its cochaperones while it is outcompeted tenfold in concentration by that of all of its clients[20]. Under stress conditions, Hsp90 expression is strongly upregulated[21]. The active Hsp90 dimer[22] proceeds through an intricate functional cycle involving ATP hydrolysis, multiple conformational rearrangements, and (transient) protein-protein interactions, whose functional roles have remained largely enigmatic despite over 40 years of Hsp90 research[23,24]. This is surprising for an ATPase as prevalent as Hsp90, and clearly sets Hsp90 apart from many other important ATPases with very well understood molecular mechanisms. In this Review, we highlight the most important '*known unknowns*' of Hsp90's molecular function, with the goal of inspiring new routes to address them and advance the field (**Fig. 1**).

Structurally, Hsp90 monomers consist of three domains (**Fig. 2A**): an N-terminal domain (NTD), a middle domain (MD), and a carboxy-terminal domain (CTD), as well as an unstructured charged linker region between the NTD and MD. The NTD contains an ATP-binding pocket and a flexible loop segment, the so-called lid, which can fold over the ATP pocket[25,26]. The charged linker modulates Hsp90 function via NTD interactions[27,28] and can influence client binding[29]. Most known client binding sites are located in the MD[2,30], while specific MD residues also participate in nucleotide binding, alongside NTD residues[31]. The CTD provides the main dimerization interface[32], and can participate in client interactions[33]. In cytosolic eukaryotic Hsp90, the CTD contains an additional recognition motif (MEEVD) for tetratricopeptide repeat (TPR)-containing cochaperones, such as Sti1/Hop and Cpr6[34]. The Hsp90 dimer undergoes large conformational rearrangements, involving conformational ensembles with several open (V-shaped), intermediate, and compacted dimer structures[34–38], as illustrated in **Fig. 1** (grey shading). Hsp90 is also assisted by cochaperones that either directly modify clients, or interact with Hsp90 to modulate client processing[39], resulting in a vast variety of transiently occurring chaperone complexes. Given this polymorphic behaviour, it has been impossible to define a single functional cycle for Hsp90, which – alongside other peculiarities discussed below – limits the current understanding of the Hsp90 functional mechanism.

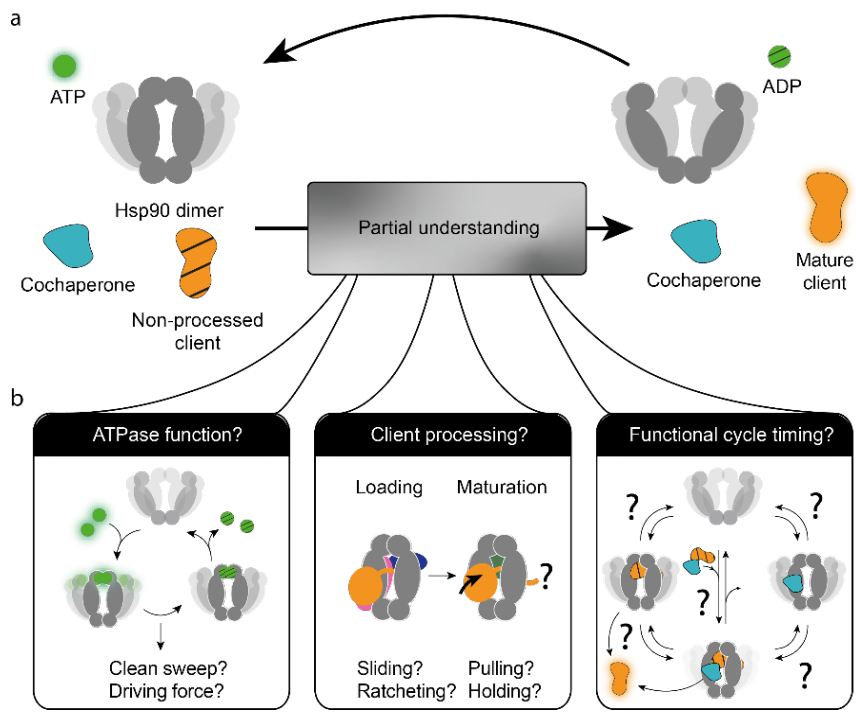

**Figure 1. Three central questions about Hsp90's functional mechanism.** (**a**) The current combined knowledge describes a general functional cycle of eukaryotic Hsp90, where Hsp90 hydrolyses ATP at a slow rate, interacts with cochaperones and clients, and (by only partially understood mechanisms) matures clients to their functional form. (**b**) Several open questions remain, as illustrated: What is the role of Hsp90's ATPase activity? How does Hsp90, alone or in cooperation with cochaperones, process its vast client base? Which is the sequence of interactions by which the Hsp90 system operates to produce functional clients? And which are reversible processes as compared to unidirectionally driven ones? Which are fast or rate-limiting steps?

**Peculiarities of the Hsp90 protein system**

For common substrate-binding ATPase proteins, such as proteases, helicases, motor proteins etc., the ATPase mechanism and function are well-defined. In stark contrast, Hsp90's function is less deterministic and (likely for that reason) less well-understood. A comparative overview of Hsp90 and a selection of well-known ATPases (**Table 1**) reveals several unique features of Hsp90.

First, Hsp90's ATP hydrolysis rate is very slow (0.1 – 1.5 min[-1])[22,40–42] – tens to hundreds of times slower than other ATPases in both presence or absence of substrate. This is a common feature of the so-called GHKL family (including the eponymous DNA *G*yrase, *H*sp, Histidine *K*inase (HisK), and Mut*L*), which share slow ATPase rates (generally 0.02 – 10/min[43–45], and DNA Gyrase with ~100/min[46]). Second, the role of ATP hydrolysis is well understood in the molecular mechanism of many standard ATPases. Contrastingly, the function of ATP hydrolysis by Hsp90 has remained unclear, and several hypotheses have been posed by the Hsp90 field, which we discuss below.

Next, the *molecular mechanisms* by which Hsp90 chaperones its clients are not well understood, since often only the *outcomes* (e.g., a functional client) are measurable, but not the molecular process leading there. This contrasts with most ATPases, where the molecular mechanism is well-defined (e.g., myosin's power stroke[47], kinesin's hand-over-hand cargo transport[48], or protein threading through protease rings for degradation[49]). Moreover, most ATPases show a specific reaction coordinate (e.g., directional translocation of an unfolded protein[49], or sliding along DNA[50]). For Hsp90, no well-defined reaction coordinate seems to exist. In fact, it interacts with a broad clientele, and with (client-specific) sets of cochaperones involved[1], as opposed to standard ATPases recognizing one specific substrate or motif. These numerous interactions occur with low affinity (up to the micromolar range[30,51,52]) and across the entire protein, not just at one specific client-binding site. While some clients bind within the cleft between both monomers (referred to as the lumen[2–4,7,53]) others bind outside of it[1,51], and cochaperones interact with all domains of Hsp90. Ultimately, Hsp90 achieves the seemingly impossible, namely, to chaperone structurally and functionally diverse clients in *client-specific* ways[54,55], which is assumed to be orchestrated via its many cochaperone interactions.

**Table 1:** Special features of Hsp90 compared to other well-known ATPase proteins: (i) the motor protein kinesin, (ii) the helicase DnaB, (iii) the protease ClpXP.

| Functional feature | Eukaryotic Hsp90 | Model ATPases |
|---|---|---|
| Mechanistic role of ATPase: | Unclear[66,79] | Understood[†] |
| ATPase rate[¥]: | Slow: 0.1 – 5 / minute[67,77,143,144] | Fast: 1 – 100 / second[145–149] |
| Substrate specificity: | Weak: diverse clients[1] | Specific[§] |
| Interaction affinities: | Low[51,52] | Medium to high[150–153] |
| ATPase functional cycle: | Non-deterministic[65,66] | Deterministic[49,145,154] |

†: (i) hand-over-hand transport[48,155]; (ii) DNA unwinding[156]; (iii) unfolded protein translocation[49].
§: (i) microtubules[155]; (ii) DNA[156]; (iii) recognition motifs[157].
¥: In presence of substrate.

Lastly, for many proteins the timing of their functional cycle is well understood: which functional states are involved (including rare, potentially rate-limiting intermediates), at what rates they occur, whether they occur sequentially or reversibly, and which of them are deterministically driven by ATP hydrolysis (**Table 1**, ATPase mechanistic role & functional cycle). This quantitative information provides a mechanistic understanding of the molecular system which aids rational drug and therapy design. As illustrated in **Fig. 1**, for Hsp90, this detailed molecular level of understanding is yet to be achieved, which may have had an impact on the relatively low success of Hsp90-targeted drugs up to now[56,57]. However, a vast body of existing biochemical and structural results about client-specific functional cycles exists and indicates possible routes forward.

## The enigmatic role of Hsp90's ATPase function

The precise functional role of the slow ATPase of Hsp90 has remained unsolved despite intense research. We first discuss the structural basis of ATP binding and the hydrolysis mechanism. Hsp90 binds ATP in a unique Bergerat fold[58]. The active site of ATP hydrolysis in yeast Hsp90 comprises five amino acid residues (**Fig. 2a**[31]): in the NTD R32, E33, and N37, and in the MD R380 and E381[25,31,35]. N37 is involved in coordinating ATP via a magnesium ion[59]. Hybrid quantum/classical (QM/MM) free-energy calculations[26], combined with large-scale atomistic molecular dynamics (MD) simulations, suggest that upon opening of the R32-E33 ion pair (**Fig. 2a**, rightmost panel) via long-range interaction and conformational switching, E33 deprotonates a water molecule in the active site, resulting in a hydroxide ion which is stabilized by R380, subsequently allowing nucleophilic attack of the bond linking the β- and γ-phosphate in the bound ATP, in which the magnesium ion coordinated by N37 pulls away electron density from the γ-phosphate to lower the hydrolysis energy barrier. After hydrolysis, R380 stabilizes the resultant cleaved inorganic phosphate[31]. The coordinated structural changes required for ATP hydrolysis are understood to be rate-limiting for Hsp90's slow ATPase cycle[25,60–62]. In other words, the probability of being in the hydrolysis competent state[63] is low due to the general flexibility of Hsp90 and, in particular, its split ATP binding site[35] composed of residues of both the NTD and MD. Last but not least, the measured ca. 3-fold higher affinity for ADP than ATP[59,64] implies a considerable product-inhibited fraction of ADP-bound Hsp90 even in the presence of excess ATP, which is expected to affect interactor binding and client maturation also *in vivo*[31].

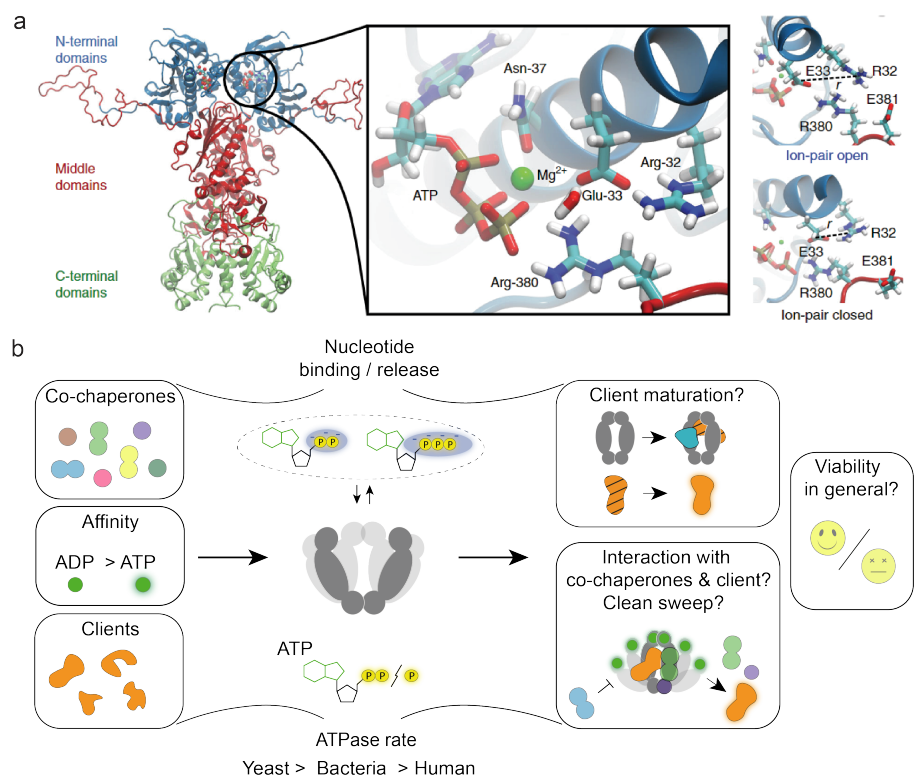

**Figure 2. The role of Hsp90's ATPase function. (a)** Structure of the yeast Hsp90 dimer (PDB ID: 2CG9)[35]. The insert obtained from an MD simulation shows the active site with bound ATP. MD simulations indicate a role of ion pair opening in the ATP-hydrolysis reaction. Figure taken from [31] (CC BY 4.0) **(b)** Visual summary of Hsp90`s ATPase rate modulation by interactors, ATPase rate differences among different species and, partially understood aspects of Hsp90's ATPase function, such as interactor dissociation. The notion that ATP hydrolysis is essential for yeast cell viability was challenged by recent evidence. Nucleotide binding and release may be sufficient to support yeast cell viability.

Interestingly, prokaryotic and eukaryotic Hsp90 show differences regarding the mechano-chemical coupling of ATP hydrolysis and large-scale structural rearrangements of the Hsp90 dimer (V-shaped opening/closing). Prokaryotic Hsp90 (subsequently referred to as HtpG) is described to function as a Brownian ratchet, where thermal fluctuations are biased by ATP-binding towards the closed state, providing a degree of directionality to its conformational cycle[38]. This has not been observed for eukaryotic Hsp90 ($Hsp90_{euk}$) where thermal fluctuations dominate even in the presence of ATP[65,66], and its conformational equilibrium is more affected by cochaperone and client interactions, as well as molecular crowding[52,63,67,68]. Nevertheless, $Hsp90_{euk}$'s nucleotide state modulates the cochaperone- and therefore client-interaction affinities. For example, ATP-binding induces NTD rotation in $Hsp90_{euk}$, affecting distal client binding sites through long-range allosteric communication, increasing the binding affinity of some (GR-LBD, MR-LBD) but not all studied clients (p53-DBD, Tau)[51,67].

A number of functional roles of Hsp90's ATPase activity have been discussed. For example, it was suggested that it facilitates the dissociation of interaction partners during GR maturation[69] (illustrated in **Fig. 2b**). Interestingly, also for MutL – another member of the GHKL class with a similar ATP binding site and hydrolysis rate – ATPase activity induces (DNA) substrate dissociation[45]. In the Hsp90-Cdc37-kinase cycle, Cdc37 inhibits Hsp90's ATPase[41,70] while clients are held in the complex[55], suggesting a role of ATP hydrolysis in regulated client dissociation – which is further supported by molecular dynamics simulations[71]. A similar case is the ATPase inhibition by p23 during GR processing[72]. While for some clients, ATP hydrolysis by Hsp90 was found to be necessary (e.g., folding of luciferase[73] and reversal of Hsp70-induced inhibition of GR activation by Hsp90[69]), in other cases, only ATP binding but not hydrolysis is required for client processing (e.g., cochaperone-independent p53 processing[74] and stabilization of the client kinase v-Src at elevated temperatures[75]). In addition, Hsp90 and its clients *mutually* affect each other and some clients alter Hsp90's ATPase rate: interestingly, ATPase stimulation (α-synuclein and ribosomal protein L2[76,77]) as well as reduction (GR-LBD client[52,67]) have been observed. Both effects were attributed to conformational stabilization of either ATPase competent or non-competent states, respectively. In this way, some clients have been suggested to modulate their own residence time on Hsp90[52]. Other clients, such as the mineralocorticoid receptor (MR)-LBD, Tau, and the p53-DNA binding domain (DBD) seem not to affect Hsp90$_{euk}$'s ATPase activity[67].

Altogether, the large variation of reported effects of Hsp90's ATPase activity remains puzzling. This is even more so, in light of the unsolved question: is Hsp90's ATPase activity indeed essential for cell viability? While, up to now, most reports maintain the notion that ATPase activity is essential for Hsp90's function and viability of the eukaryotic cell[40,78], recent evidence challenges this view: a hydrolysis-dead yeast Hsp90 E33A mutant was unexpectedly found to still support yeast cell growth[66]. Along with a recent study covering several Hsp90 orthologs with this mutation[79], these findings imply that the ability to adopt different conformations is enough for some functions of Hsp90, and that nucleotide association and dissociation may be sufficient for regulation of Hsp90's conformational equilibrium, while hydrolysis may play a different role. In summary, the large body of experimental evidence for (i) the mutual effects of cochaperones and clients on Hsp90's ATPase rate and for (ii) the role of ATP hydrolysis in client processing and cell viability, raise new questions and we discuss possible experiments to address them further below.

## Client processing by Hsp90

Client processing by $Hsp90_{euk}$ can be classified into refolding[69,80], sorting for degradation[51,76,81], ligand binding regulation[69,82,83], re-/de-activation[14,15], and regulation of assemblies[84–86]. Hsp90's diverse clients differ both structurally and functionally[11,87,88] and, related to that, Hsp90 does not have a well-defined client binding site, which is in contrast to other chaperones, such as Hsp70[89]. Instead, Hsp90 clients form numerous low-affinity contacts through hydrophobic interactions and some client-specific charge-charge interactions[72]. Certain clients (e.g., GR and kinases) primarily bind to Hsp90's lumen formed by both MDs, while intrinsically disordered proteins (IDPs, e.g.: Tau, misfolded transthyretin, etc.) bind interfaces extending from the MD to the NTD[90,91]. Despite the breadth of clients, Hsp90 is understood to exhibit client-specific chaperoning action, presumably through specific sets of cochaperones[2–7,92]. Indeed, only certain cochaperones function as integral components of the $Hsp90_{euk}$ system, while others are client specific and provide Hsp90 with the necessary plasticity[93,94] (**Fig. 3a**), for example, through specific client recruitment[11] or by modulating the time for which a client is bound to Hsp90[93,95–97].

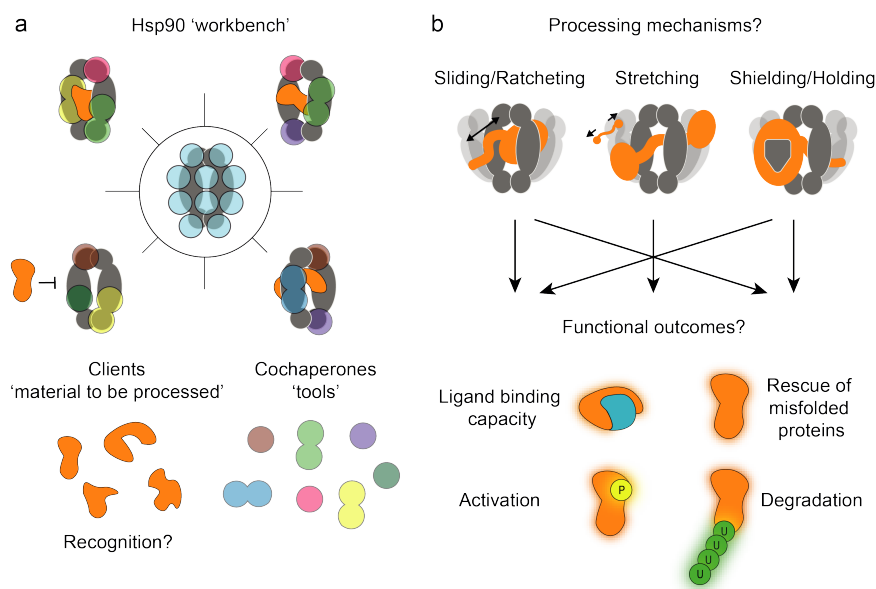

**Figure 3: Convergent model of Hsp90's functional mechanism. (a)** Hsp90 can be regarded as a workbench, bringing clients and cochaperones (coloured tools) into close proximity via interaction sites spread over the entire protein (blue spots). However, many details of recognition of the numerous clients by Hsp90 and its cochaperones remain elusive. (**b**) Various client processing modes are being discussed based on growing experimental evidence, such as sliding, ratcheting, stretching, shielding, and holding. These processing modes seem to be client-specific, and they may offer distinct functional outcomes (e.g., misfolding rescue, or targeting for or protection from degradation).

Exciting molecular-level details of client processing have recently been revealed for the GR, Hop/Sti, p23 system in 3D structures of these multi-component complexes, solved by the Agard lab[98,99] with cryo-EM. They represent two snapshots along the complex client processing trajectory, termed the loading complex[3] and the maturation complex[2]. Together, they suggest a sliding or ratcheting movement of the client GR through the Hsp90 lumen. Likewise, three kinase clients (BRAF$_{V600E}$[7], Cdk4[4], RAF1[8]) and the aryl hydrocarbon receptor[6] also thread through the Hsp90 lumen. Such molecular clamping by Hsp90 was suggested to keep the domains of multidomain clients separated, thus facilitating their independent folding[2] and preventing inter-domain misfolding (**Fig. 3b**). Destabilized kinases are thought to be held by Hsp90 to facilitate their reactivation[55], after local kinase motions poise the kinase for Hsp90 processing[100]. During the processing of Tau, which binds to open Hsp90$_{euk}$ in numerous conformations[51], Hsp90 appears to act as a holdase, preventing the formation of toxic oligomeric species until Tau is targeted for degradation[91]. Overall, Hsp90's structural plasticity allows clients and cochaperones to individually orchestrate client-processing mechanisms, making it a flexible master chaperone.

The energetic driving forces for Hsp90's diverse molecular modes of action – folding, holding, stretching, etc. – result from different underlying kinetics of inter- and intra-molecular interactions. For example, the balance between client holding versus folding depends on the (forward and reverse) kinetic rate constants of the two processes: if the client-folding step is slower than client (re-)binding to the chaperone, the holdase function is kinetically favoured and folding is rate-limiting for that particular client[101]. The sequential refolding of individual structural elements of the client can decrease their local affinity for Hsp90, enabling gradual dissociation from Hsp90. Transient interactions are prerequisite for this dynamic model. Thermodynamically, transient interactions can result in avidity effects where multiple low-affinity contacts add up, resulting in high affinities with down to nanomolar dissociation constants ($K_d$)[102]. While such strong avidity is characterized by large interfaces ($> \sim 1500$ Å$^2$), the low affinities of Hsp90 to its clients (micromolar $K_d$) [51,52] likely result from much smaller interfaces.

For folding processes to occur, specific energy barriers need to be overcome. While small conformational changes have low energy barriers and occur spontaneously in thermal equilibrium, large rearrangements with higher energy barriers are not accessible by thermal energy alone (1 $k_BT \approx$ 4.1x10$^{-21}$ J $\triangleq$ 0.6 kcal/mol). In other words, although folding is an exothermic process (cf. Anfinsen's dogma[103]), energy barriers can kinetically trap un- or misfolded states and external energy is often required for efficient folding. This can be binding energy from molecular interactions (including cochaperone binding), ATP hydrolysis energy, or potential energy changes provided by covalent post-translational modifications (PTMs). Notably, for its client-specific function, Hsp90 exploits all of these modes of action in one unified system, and first efforts have been made to directly quantify the energies involved[104]. Overall, Hsp90$_{euk}$ has emerged as a highly multi-functional workbench that uses cochaperones as tools for the customized processing of its diverse clients. Many proposed models for client processing are yet to be experimentally proven. New routes to address this are discussed below and in **Box 1**.

## The timing of Hsp90's functional cycle

The precise temporal progression through Hsp90's functional cycle is a matter of ongoing research (**Fig. 4**), complicated (amongst others) by Hsp90's client-specific behaviour, leading to many different functional cycles proposed in the literature[4–6,69]. We focus here on two of the best-studied systems, the Hsp70-Hsp90 cascade and the kinase cycle, which both progress through three main phases: (i) client loading/recruitment, leading to (ii) client processing and maturation, followed by (iii) client release. Both examples show cochaperone-dependent client loading, but we note that several clients bind Hsp90 without cochaperone involvement, for example the IDPs Tau[51] and alpha-synuclein[105]. Also, due to the generally transient low-affinity interactions, numerous intermediate complexes and their interconversion kinetics are yet to be revealed and quantified. Therefore, as direct real-time observations are still lacking, the depicted cycles represent early models from which neither a strict order of events nor their timing can be inferred.

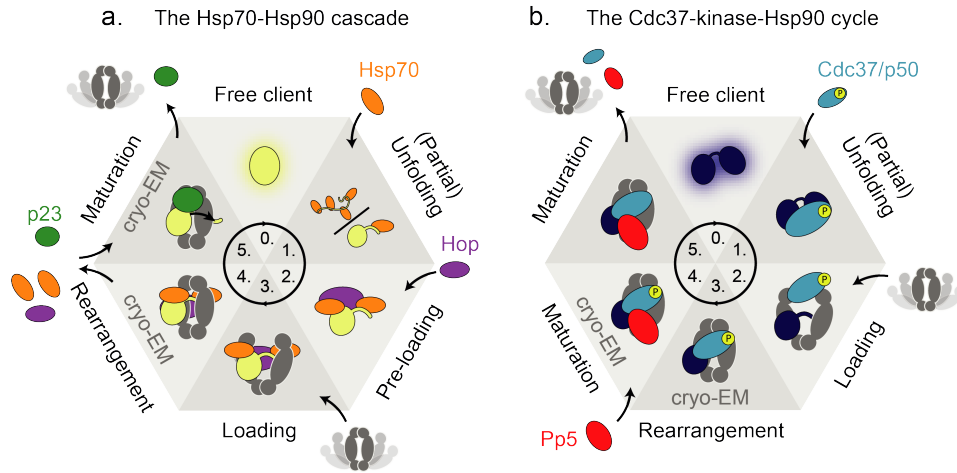

**Figure 4: Hsp90's functional cycle timing illustrated by two literature examples.** Schematic representation of the two discussed functional cycles of eukaryotic Hsp90, illustrated using six states each. Cryo-EM structures of two of the states exist, as indicated. Many transient interactions and intermediate complexes limit the current understanding of the order and timing of events, as well as their directionality or reversibility. (**a**) The Hsp70-Hsp90 cascade[2,3,69,73,80] and (**b**) the Cdc37-kinase-Hsp90 cycle[4,11]. Both cycles have in common that the client is (partially) unfolded prior to recruitment to Hsp90 in an intermediate loading complex. Next, by unknown mechanisms, clients are processed by Hsp90 through at least one intermediate maturation state. It is not completely understood, among other details, when ATP binding or hydrolysis by Hsp90 occurs, and how clients are released from these complexes.

The Hsp70-Hsp90 cascade (**Fig. 4a**) was found to process the transcription factors GR[2,3,69,80] and p53[106], as well as Argonaute which regulates transcription through RNA silencing[107]. Clients are prepared for Hsp90 interaction through (partial) unfolding by Hsp70 via hydrophobic interactions[108]. Two recently resolved multi-partite intermediate states, the loading and the maturation complex[2,3], provide much awaited molecular-level structural details. The loading complex, Hsp70-Bridge-Client-Hsp90[3], contains a 'bridging' protein (e.g., Hop[80], NudC[10], Tomm34[109]) that facilitates client transfer from Hsp70 to Hsp90. Interestingly, Hop was found not to be required *in vivo*[110] which may reflect redundancy among bridging proteins. Another bridging protein, NudC, was found to directly recruit Hsp40-bound clients to Hsp90[10]. Following recruitment, the partially unfolded client is clamped within the closed Hsp90 dimer lumen, and whether this is a directional or reversible step still needs to be determined by time-resolved techniques. Next, the Hsp90-client-p23 maturation complex[2] is formed after dissociation of the bridging protein and Hsp70, where dissociation involves either nucleotide exchange or Hsp90 ATPase activity[69]. p23 binds and stabilizes Hsp90's closed state[111] during GR processing, while it is not required for luciferase processing[80]. Interestingly, a single-molecule FRET study showed directionality of an Hsp90/p23 system conferred by ATP binding and hydrolysis even in the absence of a client protein[68], implying that cochaperones can steer the Hsp90 cycle. Additionally, p23 binding was found to relieve a Hop-inhibited Hsp70-Hsp40-Hop-GRLBD-Hsp90 complex and maximize GR hormone binding recovery beyond the levels achieved by the binding of either Hop or p23 alone[112], further indicating chaperone-endowed directionality in the Hsp70-Hsp90 cascade. The position of GR captured in the Hsp90 lumen shifts by several amino acids between the 3D-resolved loading and the maturation states, implying a sliding or ratcheting movement[2]. Although not all molecular details are known yet (e.g., on client release), the Hsp70-Hsp90 cascade appears to enable client activation through assisted folding[69,80].

Another functional cycle has been described for kinase-specific processing[4,11] (**Fig. 4b).** Here, Cdc37 plays the client-recruiting role, where the loading complex consists of Cdc37-Client-Hsp90. Cdc37 stabilizes a partially unfolded kinase state[100,113], in which the kinase N and C domain are separated[4], and thereby primes the kinase client for subsequent interaction with residues in the Hsp90 dimer lumen[55]. Cdc37 can bind Hsp90 in its open state, inhibiting Hsp90's ATPase through interaction with its ATP lid[70]. In this loading complex, phosphorylation of serine 13 on Cdc37 (Cdc37- pSer13) by casein kinase 2[114] is critical for kinase activation by Hsp90[114,115], and stabilizes the kinase-bound conformation[4]. In general, such post-translational modifications can affect the conformational dynamics, ATPase activity, client and cochaperone interactions, and other functions of Hsp90[33,116]. Phosphatases and kinases regulate client maturation by changing the phosphorylation status of Hsp90[117,118]: certain phosphorylations of Hsp90 represent allosteric regulation points[119], others promote various interactions during the functional cycle[120]. Cochaperone function is also regulated by phosphorylation, for example Hop phosphorylation at residue Y354 regulates its binding to Hsp70/90[121].

The maturation complex is obtained by binding of Protein phosphatase 5 (Pp5), whereafter Pp5 dephosphorylates the kinase clients[7] as well as Cdc37-pSer13 inducing client release[122,123]. As Cdc37's pSer13 residue is buried in the closed, ATP-bound Hsp90 dimer[7], the Pp5 dephosphorylation of Cdc37 likely happens upon Hsp90 transitioning from a compact ATP-bound state to a post-hydrolysis state, which is more flexible than the former[124]. The Hsp90-Cdc37-kinase system has been described to process clients by "factory resetting" of phosphorylations[7], as well as physical stabilization or shielding of clients[55].

The timing of functional cycles is known to be crucial for biomolecular systems and their precise regulation. Prominent examples include: the pausing mechanism of RNA polymerase[125,126], ribosome rotation, stalling, and frameshifting during translation[127], processive protein threading by proteases[49], etc. For Hsp90, this level of mechanistic understanding has not yet been reached. Up to now, only some of the state transitions of the functional cycle could be observed experimentally, leaving a range of questions unanswered: When and for what is ATP binding required? When does ATP hydrolysis take place? Is ATP binding or hydrolysis or rather cochaperone phosphorylation dominant in providing directionality – and hence processivity – to a given Hsp90 cycle? *In cellulo*, do cochaperones interact and dissociate from Hsp90 in a given sequence or randomly? What determines which processing mechanism is employed? What is the timing of specific state transitions during client processing? Clearly, the time-resolved direct observation of Hsp90's functional cycles (see below) would hold tremendous information about cycle timing, off-target waiting states, the client-specific differences, and the energetic driving forces of Hsp90's functional cycle.

**Single-molecule observations complement the mechanistic understanding of Hsp90**

Single-molecule methods have provided unique insight into Hsp90, and their abilities keep growing. By providing direct observations of individual proteins and protein complexes, they complement ensemble techniques (such as bulk enzymatic/activity assays, nuclear magnetic resonance, mass spectrometry, X-ray crystallography, etc.). Amongst others, single-molecule methods can uncover conformational heterogeneity normally hidden in ensemble averages, reveal reversible processes, elucidate the multiple reaction rates underlying the rate-limiting one, and quantitate the energies that drive molecular mechanisms. Amongst others, single-molecule Förster resonance energy transfer (smFRET)[128] and force spectroscopies[129] are popular techniques that provide time-resolved access to protein functional determinants, such as intra- and inter-molecular dynamics[128], molecular forces[129], enzymatic reactions[130], and movement[48] at single-molecule resolution.

Specifically, using smFRET, it was revealed that prokaryotic and eukaryotic Hsp90 are distinct regarding their conformational rearrangements, which are strongly ATP-hydrolysis dependent in the prokaryotic[38] but not eukaryotic[37,131] homologue. Next, the conferral of directionality by ATP binding or hydrolysis in an Hsp90/p23 system[68] was uncovered by smFRET, and also how Hsp90's conformational kinetics are regulated by environmental conditions (e.g., the presence of point mutations, crowding agents, or cochaperones)[63]. In addition, it could be shown that Hsp90 inhibitors have little effect on Hsp90's conformational dynamics[132]. Furthermore, smFRET revealed that Grp94 (a human Hsp90 isoform confined to the endoplasmic reticulum) exhibits two closed states, which was explained to arise from sequential ATP hydrolysis by the Hsp90 protomers[133]. Also, smFRET elucidated that the yeast Hsp90 charged linker reduces the probability of the N-terminal dimerization conformation of the Hsp90 dimer as compared to Hsp90 mutants lacking charged residues in the linker[28]. Moreover, smFRET showed that the p53 DNA-binding domain exhibits considerable dynamics in the presence of all required chaperoning components (i.e., Hsp70, Hsp40, ATP, HOP, Bag1, and Hsp90), suggesting that p53 conformation is constantly remodelled by Hsp70/Hsp90[106]. In an integrative study, building on structural information, more than 100 FRET pairs on the Hsp90 dimer were used to disentangle local and global dynamics of the Hsp90 dimer, and it was found that the model client Δ131Δ only affects open state dynamics and does not affect open-closed interconversion dynamics[134]. Another study connecting smFRET and molecular dynamics observations suggests that ATP hydrolysis leads to a conformational state in which the client binding site in the Hsp90 dimer lumen becomes constricted[135]. A different fluorescence technique, single-molecule photo-induced electron transfer fluorescence correlation spectroscopy (PET-FCS) showed the degree of cooperative motion of Hsp90 domains as well as specific regions during various stages of Hsp90's catalytic cycle[25,60]. Complementary to fluorescence, single-molecule force spectroscopy studies using optical tweezers have revealed the mechanism by which Hsp90 (itself) folds[136], and elucidated the functional role of its flexible charged linker region[137]. Similar experiments have been used to perform a detailed comparison of Hsp90 orthologs, where large differences in the flexibility of the charged linker regions were identified[138]. Also, Hsp90 isoforms were investigated by force spectroscopy, where despite a very high sequence similarity, differences in stability, refolding capacity and their conformational cycles were observed[139]. Lastly, optical tweezer experiments revealed details of the ATP dependence of Hsp90's dimerization dynamics[140], of Hsp70-mediated client unfolding[112], and the local compaction of the model client luciferase in an ATP-dependent manner[104]. Altogether, by building on a large body of existing biochemical work, single-molecule techniques offer valuable complementary experiments that provided otherwise inaccessible insights through direct observations of single proteins at work.

Looking ahead, the future is bright for new approaches applied to Hsp90 – in part due to recent technological advances. For example, Hsp90's weak interactions with clients and cochaperones posed a major challenge of single-molecule experiments in the past. This regime can now be studied using nanophotonic approaches such as zero-mode waveguides (ZMW) – which overcome previous detection limitations for single-molecule fluorescence spectroscopy[127,141]. Alternatively, tethered constructs, where the local concentration of a cochaperone, for example, is enhanced by connecting it to Hsp90 through an unstructured amino acid linker, are being used for both fluorescence spectroscopy and force-based measurements[142]. In fact, many of the urgent questions discussed herein could be clarified with the help of single-molecule experiments, and, in **Box 1**, we provide a non-exhaustive list with specific examples.

**Box 1: New opportunities to elucidate Hsp90's enigmas with single-molecule resolution**

(i) The order of interactions during Hsp90's functional cycle can be observed in real-time in single-molecule fluorescence experiments, by recording the consecutive binding and dissociation of fluorescently labelled interactors to surface-tethered Hsp90. Zero-mode waveguides can be used to achieve single-molecule resolution at physiological concentrations.

(ii) Questions about Hsp90's ATPase and how it is coupled to Hsp90's conformational state can be elucidated by investigating the differences between prokaryotic and eukaryotic Hsp90 in this regard. smFRET and force spectroscopy can observe these differences in conformational dynamics and reveal the energies involved.

(iii) Client processing by Hsp90 can be observed at the single-molecule level in real-time, e.g., to probe the proposed sliding hypothesis for GR processing[2]. Such nanometer distance changes induced by sliding or ratcheting can be observed by smFRET within one chaperone complex. Force spectroscopy can also be used to detect distance changes associated with client-remodelling.

(iv) Functional cycle timing, the order of individual events, and whether they occur directionally or reversibly can be observed by single-molecule fluorescence and dye-labelled interactors (e.g., Hsp90, client, Hsp70, p23, etc.) at physiological concentrations (using ZMW). The impact of clients and cochaperones on the conformational state of Hsp90 can also be probed by using force spectroscopy experiments combined with microfluidics.

# Summary

In summary, the central chaperone Hsp90 is a peculiar ATPase differing drastically from other well-known ATPases, and it has remained surprisingly enigmatic. We highlighted three fundamental unanswered questions on Hsp90. *First*, what is the purpose of Hsp90's ATPase activity? It has been revealed that Hsp90 can show widespread allostery in its hydrolysis mechanism. Additionally, Hsp90's ATP hydrolysis was proposed to induce dissociation of interaction partners, but *in cellulo*, it was found that nucleotide exchange (not hydrolysis) was sufficient for several Hsp90 chaperoning functions. *Second*, what are the molecular actions by which the Hsp90 system processes its clients? First molecular-level insights from cryo-EM structures provided static evidence for dynamic rearrangements, such as sliding, stretching, and holding – which represent valuable hypotheses to test with time-resolved observations, e.g.: using single-molecule techniques. *Third,* what are the kinetics of Hsp90's functional cycle, or rather its multiple client-specific cycles? I.e., what is the timing, order and interdependence of events that describe this multi-component system? With first molecular structures elucidating parts of the functional cycle plus ample biochemical results on the interaction partners involved, it is now possible to set up experiments to directly observe these functional cycles by following a single multipartite complex over time.

Altogether, mechanistic insight on the multifaceted chaperone function of Hsp90 is urgently needed to inform and accelerate biomedical and pharmaceutical advances, targeting several conditions ranging from cancer to neurodegenerative diseases and more. Compared to other intensely studied protein systems, the understanding of the Hsp90 chaperone lags behind, because its transient multipartite complexes have been challenging to study in molecular detail. Nevertheless, a large body of biochemical knowledge has accumulated over the years, 3D structural information on the transient Hsp90 complexes is growing, and recent technological advances enable now direct observation of Hsp90 at work. Equipped with these, the time is better than ever to address and uncover the remaining mechanistic known unknowns of the Hsp90 chaperone.

**Acknowledgements**

We thank Prof. Johannes Buchner for a critical reading of the manuscript and for his highly insightful comments and support.  We regret not being able to cover all Hsp90 work related to our Review topic, and we acknowledge the work of all researchers working in the broad Hsp90 field. KT is supported by an MSCA-IF (NOTE, grant agreement number 101028366) and KT and SS are supported by the NWO grant (Solve90, grant agreement number OCENW.M.22.092).